\documentclass[aps,prd,showpacs,amsmath,amssymb]{revtex4}
\usepackage{graphicx}
\usepackage{dcolumn}
\usepackage{bm}
\newcommand{\be}{\begin{equation}}
\newcommand{\ee}{\end{equation}}
\newcommand{\ba}{\begin{eqnarray}}
\newcommand{\ea}{\end{eqnarray}}
\def\lsim{\lower.5ex\hbox{$\; \buildrel < \over \sim \;$}}
\def\gsim{\lower.5ex\hbox{$\; \buildrel > \over \sim \;$}}
\def\ch{\lower-0.55ex\hbox{--}\kern-0.55em{\lower0.15ex\hbox{$h$}}}
\def\lh{\lower-0.55ex\hbox{--}\kern-0.55em{\lower0.15ex\hbox{$\lambda$}}}

\begin{document}

\title{Separation of Dirac equation in the $3+1$ dimensional constant
curvature black hole background and its solution}

\author{Banibrata Mukhopadhyay}\email{bm@physics.iisc.ernet.in}
\affiliation{Department of Physics, Indian Institute of Science,
Bangalore-560012, India}
\author{Kaushik Ghosh}
\affiliation{Department of Engineering Sciences, Haldia Institute of Technology,
Haldia, East Midnapore-721 657, West Bengal, India}



\begin{abstract}

The behavior of spin-half
particles is discussed in the $3 + 1$-dimensional constant curvature
black hole (CCBH) spacetime. We use Schwarzschild-like coordinates,
valid outside the black hole event horizon.
The constant time surfaces corresponding to the
time-like Killing vector are degenerate at the black
hole event horizon and also along an axis. We write down the
Dirac equation in this spacetime using Newman-Penrose formalism
which is not easily separable unlike that in the Kerr metric.
However, with a particular choice of basis system the equation is separable
and we obtain the solutions. We discuss the structural difference in
the Dirac equation in the CCBH spacetime with that in the Kerr geometry,
due to difference in the corresponding spacetime metric, resulting 
complexity arised in separation in the earlier case.

\end{abstract}


\maketitle


\section{Introduction}

A new type of black hole solution has been found by
Ba{\~n}ados \cite{B1}. This solution results from an identification
of space-time points in the anti-de Sitter space and represents a
higher dimensional generalization of the $2+1$ dimensional BTZ black
hole \cite{B2}. This is a constant curvature black hole (CCBH) with a negative
cosmological constant. The corresponding spacetime geometry is $R^3 \times S^1$.
Unlike the BTZ black hole, the $3+1$ dimensional black
hole does not have rotating solution and is dynamic.
However Schwarzschild-like coordinates, valid outside
the horizon, can be found which covers a sub-manifold
of the space time. 

The scalar field solution in the $3+1$ dimensional CCBH
was discussed in \cite{K1} while studying their thermodynamic behavior. 
On the other hand, the spinor field solution was studied extensively
in black hole spacetimes over last three decades by various
authors including one of the present ones (e.g. \cite{teu,chandra76,page76,chandra83,guv90,mc99,mc,md,m00}).
These authors discussed in detail the separability of the Dirac equation
in black hole spacetimes and solutions. 
In the present article we study the behavior
of a spin-half field in the $3+1$ dimensional CCBH spacetime.
We use Newman-Penrose formalism to establish the Dirac equation.
It is found that in this background the Dirac equation is not easily separable, unlike the
situation in the Kerr geometry, unless we change the basis appropriately. This is,
as we discuss in the following sections in detail, due to very nature of the background
geometry which is, by definition, quite different from conventional black hole spacetimes.

The paper is organized as follows. In the next section we recall the $3+1$ CCBH metric
and describe its properties briefly. In \S III we calculate the basis vectors of null
tetrad in this spacetime and write down the Dirac equation following the Newman-Penrose 
formalism. We show that the set of equations is separable into radial and angular
part by defining new spinors obtained with a suitable combination of original 
spinors. Subsequently we discuss the solutions in \S IV. The section V is devoted
in obtaining the effective potentials of spinors in the CCBH spacetime and comparing
with that in the Kerr geometry. We show that due to difference in structure of
spacetime metric the Dirac equation in the CCBH spacetime appears different than 
that in the Kerr geometry hindering its separation in regular basis where particles are
expressible as {\it pure} up and/or down spinors. Finally, we summarize our results in \S VI.

\section{Description of the $3+1$ CCBH metric}

The anti-de Sitter spacetime in 3+1 dimension is defined as the
universal covering space of the hypersurface
\begin{equation}
-x^{2}_{0} + x^{2}_{1} + x^{2}_{2} + x^{2}_{3} - x^{2}_{4} = - l^{2},
\end{equation}
where $l^2=-1/\Lambda$, when $\Lambda$ is the cosmological constant.
The 3+1 dimensional CCBH is obtained by making identifications
in this space using an one dimensional subgroup
of its isometry group SO(2,3). In the Kruskal
coordinates the metric for the 3+1 dimensional CCBH is given by \cite
{B1},
\be
ds^{2} = {l^{2}{(r + r_h)^2}\over{r_h}^2}{dy^{\alpha}dy^{\beta}}
{\eta_{\alpha\beta}} + r^{2}d{\phi}^{2}
\ee
where $r$ is given by
\be
r = r_h{(1 + y^2)\over(1 - y^2)}
\ee
and $y^2 = {y^{\alpha}y^{\beta}}{\eta_{\alpha\beta}}$
[${\eta_{\alpha\beta}}$ = diag(-1,1,1)], $r_h$ is the black hole horizon. The
coordinate ranges are $-\infty<y^\alpha<\infty$ and
$0\le{\phi}<2{\pi}$.

We introduce the local Schwarzschild-like coordinates $(t,r,\theta)$
in the 2+1 dimensional hyperplane $y^\alpha$ as
\ba
 y^0 & = & {\alpha (r)}\cos{\theta}\sinh({{r_h}t\over l})\nonumber\\
y^1 & = & {\alpha (r)}\cos{\theta}\cosh({{r_h}t\over l})\nonumber\\
 y^2 & = & {\alpha (r)}\sin{\theta}
\ea
where $\alpha (r) = [{(r - r_h)\over(r + r_h)}]^{1\over2}$.

In these coordinates the metric becomes
\be
ds^{2} = {l^{4}f^{2}(r)\over r^{2}_h}[d{\theta}^{2} -
\cos^{2}{\theta}(dt/l)^{2}]
+ {dr^{2}\over f^{2}(r)} + r^{2}d{\phi}^{2}
\ee
where ${f(r) = ({{{r^{2}} - {r^{2}_h}}\over{l^{2}}})^{1\over2}}$.
These coordinates are valid outside the horizon ($r>r_{h}$)
for $-{\pi \over 2}<{\theta}<{\pi \over 2}$
and $0\le{\phi}<2{\pi}$. However in these coordinates only part of the
space is covered $(-1<{y_2}<1)$.
It is clear that the foliation becomes degenerate
along the direction  ${\theta} = -{\pi\over2}$ and ${\theta} = {\pi\over2}$.

\section{Dirac equation in the CCBH metric}

In this section
we will use the Newman-Penrose formalism \cite{np} to write the Dirac equation
in the above mentioned constant curvature black hole
spacetime.
The Dirac equation in Newman-Penrose formalism can be written as \cite{chandra76}
\be
\sigma_{AB'}^\mu{D_\mu}P^A+i\mu_p{\bar Q}^{C'}\epsilon_{C'B'}=0,
\ee

\be
\sigma_{AB'}^\mu{D_\mu}Q^A+i\mu_p{\bar P}^{C'}\epsilon_{C'B'}=0,
\ee
where, for any vector $X_i$, according to the spinor formalism 
$\sigma_{AB'}^iX_i=X_{AB'}$, $\epsilon_{C'B'}$ is the twodimensional Levi-Civita tensor; 
$A,B=0,1$ and
\be
D_\mu P^A=\partial\mu P^A+\Gamma^A_{\mu\nu} P^\nu. 
\ee

Here we introduce a null tetrad $(\vec{l}, \vec{n}, \vec{m}, \vec{\bar{m}})$
to satisfy orthogonality relations, $\vec{l}{\bf .}\vec{n}=1$, 
$\vec{m}{\bf .}\vec{\bar{m}}=-1$ and $\vec{l}{\bf .}\vec{m}=\vec{n}{\bf .}\vec{m}=
\vec{l}{\bf .}\vec{\bar{m}}=\vec{n}{\bf .}\vec{\bar{m}}=0$ following 
Newman \& Penrose \cite{np}. $2^{\frac{1}{2}}\mu_p$ is the mass of the Dirac particle. 
In terms of this new basis, in Newman-Penrose formalism, the 
Pauli matrices can be written as \cite{chandra83}

\be
\sigma_{AB'}^\mu=\frac{1}{\sqrt{2}}\left(
\begin{array}{cr} l^{\mu} & m^{\mu}\\{\bar{m}}^{\mu} 
& n^{\mu}\end{array}\right).
\ee

%
%
%
Now following previous works \cite{chandra83,m00},
writing various spin coefficients by their named symbols \cite{chandra83}, 
and choosing 
$$
P^0=F_1, P^1=F_2, {\bar Q}^{1'}=G_1, {\bar Q}^{0}=-G_2
$$ 
we obtain 
\be
l^{\mu}\partial_{\mu}F_1+{\bar m}^{\mu}\partial_\mu F_2+
(\epsilon-\tilde{\rho}){F_1}+(\pi-\alpha){F_2}=i{\mu_p}G_1,
\ee

\be
m^{\mu}\partial_{\mu} F_1+n^{\mu}\partial_\mu F_2+
(\mu-\gamma){F_2}+(\beta-\tau){F_1}=i{\mu_p}G_2,
\ee

\be
l^{\mu}\partial_{\mu}G_2-m^{\mu}\partial_\mu G_1+
({\epsilon}^*-\tilde{{\rho}}^*){G_2}-({\pi}^*-{\alpha}^*){G_1}=i{\mu_p}F_2,
\ee

\be
n^{\mu}\partial_{\mu}G_1-{\bar m}^{\mu}\partial_\mu G_2+
({\mu}^*-{\gamma}^*){G_1}-({\beta}^*-{\tau}^*){G_2}=i{\mu_p}F_1.
\ee
These are the Dirac equations in Newman-Penrose formalism in curved
space-time in absence of electromagnetic interaction. The corresponding equations 
in presence of electromagnetic interaction can be found in earlier works 
(e.g. \cite{pr,md,hn}). 

Now we calculate the basis vectors of null tetrad in terms of elements of the CCBH metric 
given as
$$
l^{\mu}=\left(\frac{r_h}{l\cos\theta f^2},1,0,0\right),
\eqno{(14a)}
$$
$$
n^{\mu}=\left(\frac{r_h}{2l\cos\theta},-\frac{f^2}{2},0,0\right), 
\eqno{(14b)}
$$
$$
m^{\mu}=\left(0,0,\frac{r_h}{\sqrt{2}l^2f},\frac{i}{\sqrt{2}r}\right),
\eqno{(14c)}
$$
$$
{\bar m}^{\mu}=\left(0,0,\frac{r_h}{\sqrt{2}l^2f},-\frac{i}{\sqrt{2}r}\right)
\eqno{(14d)}
$$
and
$$
l_{\mu}=\left(\frac{l\cos\theta}{r_h},-\frac{1}{f^2},0,0\right),
\eqno{(15a)}
$$
$$
n_{\mu}=\left(\frac{f^2 l \cos\theta}{2r_h},\frac{1}{2},0,0\right),
\eqno{(15b)}
$$
$$
m_{\mu}=\left(0,0,-\frac{fl^2}{\sqrt{2}r_h},-\frac{ir}{\sqrt{2}}\right),
\eqno{(15c)}
$$
$$
{\bar m}_{\mu}=\left(0,0,-\frac{fl^2}{\sqrt{2}r_h},\frac{ir}{\sqrt{2}}\right).
\eqno{(15d)}
$$
We also calculate the various spin coefficients as
$$
\tilde{\rho}=-\frac{1}{2}\left(\frac{\partial_r f}{f}+\frac{1}{r}\right),\hskip0.5cm \beta=0,
\hskip0.5cm \pi=-\frac{r_h \tan\theta}{2\sqrt{2}l^2f},\hskip0.5cm \epsilon=0,
$$
$$
\tau=\frac{r_h \tan\theta}{2\sqrt{2}l^2f},\hskip0.5cm 
\mu=-\frac{f^2}{4}\left(\frac{\partial_r f}{f}+\frac{1}{r}\right),
\hskip0.5cm \gamma=\frac{f \partial_r f}{2},\hskip0.5cm \alpha=0.
\eqno{(16)}
$$
Now we consider the spin-$\frac{1}{2}$ wave function as the form of 
$e^{i({\sigma}t+m{\phi})}F(r,\theta)$, where $\sigma$ is the frequency of
the incoming wave and $m$ is the azimuthal quantum number.

Thus using eqns. (14), (15), and (16), eqns. (10) to (13) reduce to
$$
\left[\frac{i\sigma r_h}{l\cos\theta f}+f\partial_r+\frac{1}{2}\left(\partial_r f+
\frac{f}{r}\right)\right]F_1+\left[\frac{r_h}{\sqrt{2}l^2}\partial_\theta-\frac{r_h 
\tan\theta}{2\sqrt{2}l^2}+\frac{mf}{\sqrt{2}r}\right]F_2=i\mu_pfG_1,
\eqno{(17a)}
$$
$$
\left[\frac{ir_h\sigma}{2l\cos\theta}-\frac{f^2}{2}\partial_r-\frac{f^2}{4}
\left(\frac{\partial_r f}{f}+\frac{1}{r}\right)-\frac{f \partial_r f}{2}\right]F_2
+\frac{1}{f}\left[\frac{r_h}{\sqrt{2}l^2}\partial_\theta-
\frac{r_h \tan\theta}{2\sqrt{2} l^2}-\frac{mf}{\sqrt{2}r}\right]F_1=i\mu_pG_2,
\eqno{(17b)}
$$
$$
\left[\frac{i\sigma r_h}{l\cos\theta f}+f\partial_r+\frac{1}{2}\left(\partial_r f+
\frac{f}{r}\right)\right]G_2-\left[\frac{r_h}{\sqrt{2}l^2}\partial_\theta-\frac{r_h
\tan\theta}{2\sqrt{2}l^2}-\frac{mf}{\sqrt{2}r}\right]G_1=i\mu_pfF_2,
\eqno{(17c)}
$$
$$
\left[\frac{ir_h\sigma}{2l\cos\theta}-\frac{f^2}{2}\partial_r-\frac{f^2}{4}
\left(\frac{\partial_r f}{f}+\frac{1}{r}\right)-\frac{f \partial_r f}{2}\right]G_1
-\frac{1}{f}\left[\frac{r_h}{\sqrt{2}l^2}\partial_\theta-
\frac{r_h \tan\theta}{2\sqrt{2} l^2}+\frac{mf}{\sqrt{2}r}\right]G_2=i\mu_pF_1.
\eqno{(17d)}
$$
Now we define 
$$
\frac{\sqrt{2}F_1}{f}=\tilde{f}_1, F_2=f_2, G_1=g_1,\,\ {\rm and}\,\, \frac{\sqrt{2}G_2}{f}
=-\tilde{g}_2,
\eqno{(18)}
$$ 
and 
$$
{\cal D}=f^2\left(\partial_r+\frac{1}{2r}\right)+\frac{3}{2}f\partial_r f,\,\,\,\,
{\cal L}=\frac{r_h}{l^2}\partial_\theta-\frac{r_h \tan\theta}{2l^2},
\eqno{(19)}
$$
and reduce eqns. (17a-d) to
$$
\left({\cal D}+\frac{i\sigma r_h}{l\cos\theta}\right)\tilde{f}_1+\left({\cal L}
+\frac{mf}{r}\right){f}_2=i\sqrt{2}\mu_pfg_1,
\eqno{(20a)}
$$
$$
\left({\cal D}-\frac{i\sigma r_h}{l\cos\theta}\right){f}_2-
\left({\cal L}-\frac{mf}{r}\right)\tilde{f}_1=i\sqrt{2}\mu_pf\tilde{g}_2,
\eqno{(20b)}
$$
$$
\left({\cal D}+\frac{i\sigma r_h}{l\cos\theta}\right)\tilde{g}_2+\left({\cal L}
-\frac{mf}{r}\right)g_1=-i\sqrt{2}\mu_p f{f}_2,
\eqno{(20c)}
$$
$$
\left({\cal D}-\frac{i\sigma r_h}{l\cos\theta}\right)g_1-
\left({\cal L}+\frac{mf}{r}\right)\tilde{g}_2=-i\sqrt{2}\mu_pf\tilde{f}_1.
\eqno{(20d)}
$$

We now follow the Chandrasekhar's procedure \cite{chandra76,chandra83} which was generalised by
Page \cite{page76}, Carter and McLenaghan \cite{cm}. The last two authors showed explicitly
that the method consists of two steps. First, one has to replace the original wave
equation by a modified but equivalent one such that the operator gets split up into
radial and angular part which commute each other. Second, making use of the explicit
form of the modified operators, one has to factorize the components of the spinor
into pairs of single variable, one of $r$ and other of $\theta$, interdependent 
functions obeying a system of ordinary differential equations involving a 
separation constant $\lambda$. Hence, the radial and angular operators also
commute with the total wave operator transformed from the original one. This
implies the existence of a conserved current associated with each solution
of the equation.

Carter and McLenaghan \cite{cm} showed, with the use of Penrose-Floyd tensor \cite{pf}, 
that occurrence of type of operators mentioned above arises in presence of an appropriate 
Killing spinor field in the spacetime under consideration. The existence of a Killing 
spinor, $K_{CD}$, was demonstrated earlier \cite{pw} 
in any type-D vacuum spacetime of a second order symmetric two-spinor satisfying 
$$
\nabla_{A^\prime(B}K_{CD)}=0,
\eqno{(21)}
$$ 
when the latin indices run from $0$ to $1$,
parenthesis indicates symmetrization over the indices within, and the primed 
spinors transform under the conjugate of the transformation from the unprimed ones
in a Lorentz transformation. The solution of above equation is a conformal Killing
spinor, since it describes a constant motion along the null geodesic. On the other
hand, constants of all the Kinnersley type-D vacuum solutions can be derived from 
separability of the null geodesic Hamilton-Jacobi equation \cite{cm}. It is therefore
appropriate to demand the symmetric two-spinor to satisfy
$$
\nabla_{A^\prime C}K^C_{B}+{\bar \nabla}_{BC^\prime}{\bar K}^{C^\prime}_{A^\prime}=0.
\eqno{(22)}
$$
This later condition was understood more clearly by the Penrose-Floyd tensor 
(see \cite{cm} and \cite{pf} for details), 
$$
f_{\mu\nu}\leftrightarrow{\bar \epsilon}_{A^\prime B^\prime}K_{AB}+\epsilon_{AB}
{\bar K}_{A^\prime B^\prime},
\eqno{(23)}
$$
satisfying 
$$
f_{\mu(\nu;\delta)}=0.
\eqno{(24)} 
$$

We find that defining $\Psi_1=\tilde{f}_1+{f}_2$, $\Psi_2=f_2-\tilde{f}_1$, 
$\Psi_3=g_1+\tilde{g}_2$ and $\Psi_4=g_1-\tilde{g}_2$ and combining eqns. (20a-d) we obtain
$$
\left({\cal D}+\frac{mf}{r}\right)\Psi_1+\left({\cal L}-\frac{i\sigma r_h}{l\cos\theta}\right)\Psi_2
=i\sqrt{2}\mu_pf\Psi_3,
\eqno{(25a)}
$$
$$
\left({\cal D}-\frac{mf}{r}\right)\Psi_2-\left({\cal L}+\frac{i\sigma r_h}{l\cos\theta}\right)\Psi_1
=-i\sqrt{2}\mu_pf\Psi_4,
\eqno{(25b)}
$$
$$
\left({\cal D}-\frac{mf}{r}\right)\Psi_3+\left({\cal L}-\frac{i\sigma r_h}{l\cos\theta}\right)\Psi_4
=-i\sqrt{2}\mu_pf\Psi_1,
\eqno{(25c)}
$$
$$
\left({\cal D}+\frac{mf}{r}\right)\Psi_4-\left({\cal L}+\frac{i\sigma r_h}{l\cos\theta}\right)\Psi_3
=i\sqrt{2}\mu_pf\Psi_2
\eqno{(25d)}
$$
what satisfy all the above conditions to separate them into
radial and angular operators allowing to follow the procedure by Chandrasekhar \cite{chandra83} 
generalized by Carter and McLenaghan \cite{cm} described above.

Now defining $\Psi_1=R_{-}(r)S_{-}(\theta)$, $\Psi_2=R_{+}(r)S_{+}(\theta)$,
$\Psi_3=R_{+}(r)S_{-}(\theta)$ and $\Psi_4=R_{-}(r)S_{+}(\theta)$ we separate
the set of equations into radial and angular parts given by
$$
\left({\cal D}+\frac{mf}{r}\right)R_{-}-i\sqrt{2}\mu_pfR_{+}=\lambda_1 R_{+};
\hskip0.2cm \left({\cal L}-\frac{i\sigma r_h}{l\cos\theta}\right)S_{+}=-\lambda_1 S_{-},
\eqno{(26a)}
$$
$$
\left({\cal D}-\frac{mf}{r}\right)R_{+}+i\sqrt{2}\mu_pfR_{-}=\lambda_2 R_{-};
\hskip0.2cm \left({\cal L}+\frac{i\sigma r_h}{l\cos\theta}\right)S_{-}=\lambda_2 S_{+},
\eqno{(26b)}
$$
$$
\left({\cal D}-\frac{mf}{r}\right)R_{+}+i\sqrt{2}\mu_pfR_{-}=\lambda_3 R_{-};
\hskip0.2cm \left({\cal L}-\frac{i\sigma r_h}{l\cos\theta}\right)S_{+}=-\lambda_3 S_{-},
\eqno{(26c)}
$$
$$
\left({\cal D}+\frac{mf}{r}\right)R_{-}-i\sqrt{2}\mu_pfR_{+}=\lambda_4 R_{+};
\hskip0.2cm \left({\cal L}+\frac{i\sigma r_h}{l\cos\theta}\right)S_{-}=\lambda_4 S_{+}.
\eqno{(26d)}
$$
From the radial part of eqns. (26a-d) it is clear that for unique solution 
$\lambda_4=\lambda_1$ and $\lambda_3=\lambda_2$. Thus further eqns. (26a-d) reduce to 
$$
\left({\cal D}+\frac{mf}{r}\right)R_{-}=\left(\lambda_1+i\sqrt{2}\mu_p f\right)R_{+};
\hskip0.2cm\left({\cal D}-\frac{mf}{r}\right)R_{+}=\left(\lambda_2-i\sqrt{2}\mu_p f\right)R_{-},
\eqno{(27a)}
$$
$$
\left({\cal D}-\frac{mf}{r}\right)R_{+}=\left(\lambda_2-i\sqrt{2}\mu_p f\right)R_{-};
\hskip0.2cm\left({\cal D}+\frac{mf}{r}\right)R_{-}=\left(\lambda_1+i\sqrt{2}\mu_p f\right)R_{+},
\eqno{(27b)}
$$
$$
\left({\cal L}-\frac{i\sigma r_h}{l\cos\theta}\right)S_{+}=-\lambda_1 S_{-};
\hskip0.2cm\left({\cal L}+\frac{i\sigma r_h}{l\cos\theta}\right)S_{-}=\lambda_2 S_{+},
\eqno{(27c)}
$$
$$
\left({\cal L}-\frac{i\sigma r_h}{l\cos\theta}\right)S_{+}=-\lambda_2 S_{-};
\hskip0.2cm\left({\cal L}+\frac{i\sigma r_h}{l\cos\theta}\right)S_{-}=\lambda_1 S_{+}.
\eqno{(27d)}
$$
If we choose $\lambda_1=\lambda_2=\lambda$ and rename 
$$
{\cal D}+\frac{mf}{r}\rightarrow {\cal D}^\dagger; \hskip0.4cm{\cal D}-\frac{mf}{r}
\rightarrow {\cal D}
\eqno{(28a)}
$$
and
$$
{\cal L}+\frac{i\sigma r_h}{l\cos\theta}\rightarrow {\cal L}^\dagger; \hskip0.4cm{\cal L}-
\frac{i\sigma r_h}{l\cos\theta}\rightarrow {\cal L},
\eqno{(28b)}
$$
then eqns. (27a-d) reduce to
$$
{\cal D}R_{+}=\left(\lambda-i\sqrt{2}\mu_p f\right)R_{-};
\hskip0.2cm{\cal D}^\dagger R_{-}=\left(\lambda+i\sqrt{2}\mu_p f\right)R_{+},
\eqno{(29)}
$$
$$
{\cal L}S_{+}=-\lambda S_{-};
\hskip0.2cm{\cal L}^\dagger S_{-}=\lambda S_{+}.
\eqno{(30)}
$$
The eqns. (29-30) are the separated set of Dirac equation, where
${\cal D}^\dagger$, ${\cal D}$ and ${\cal L}^\dagger$, ${\cal L}$ act as creation 
and annihilation operators for radial and angular functions respectively.
For comparison, one may check the corresponding equations in the Kerr geometry
\cite{chandra83,mc} given below in eqns. (47) and (48).

\section{Solution for the angular and radial Dirac equation}

The angular equation for ${S_{+}}$ is given by
$$
{{\partial_\theta}^2}{S_{+}} -
\tan\theta\,{\partial}_{\theta}{S_{+}} +
{\big[ \nu (\nu + 1) - 
{{n^2 - 2n\mu \sin\theta + {\mu}^2}\over
  {{\cos^2}\theta}}\big]}{S_{+}}
=0
\eqno{(31)}
$$
where ${n = 1/2}$, ${\mu = -i\sigma l}$
and $\nu (\nu + 1) = {{l^4 {\lambda}^2}\over {r_h}^2} - 1/4$.
Following previous works \cite{CV,K2}, the solution of eqn. (31) is given by 
$$
{S_{+}}(x) \rightarrow {\big({{1 - x}\over 2}\big)^{\frac{n - \mu}{2}}}
{\big({{1 + x}\over 2}\big)^{\frac{n + \mu}{2}}}
F(a,b,c;{{1 -x}\over 2})
\eqno{(32)}
$$
where

$$
a = -\nu + {1\over2}[(n - \mu) + (n + \mu)],
$$
$$
b = \nu + 1 + {1\over2}[(n - \mu) + (n + \mu)],
$$
$$
c = (n - \mu) + 1,
$$
$$
x=\cos\theta.
\eqno{(33)}
$$
This solution is well-behaved throughout the interval
${-{\pi \over 2}}<{\theta}<{\pi \over 2}$ \cite{gr}, the range
of $\theta$ where the coordinate system is valid outside of the black hole.
The solution for $S_{-}$ is similar to $S_{+}$ with
$\mu$ replaced by $-\mu$. 

The radial equation for $R_{+}$ is given by,
$$
{f^2 (r)} {\partial_r (f^2 \partial_r R_{+})} +
{f^2 (r)} {\partial_r ([g(r) - h(r)]R_{+})} +
{f^2 (r)} [g(r) + h(r)]{\partial_r {R_{+}}} + [g^2 (r) - h^2 (r)]R_{+} 
$$
$$
-(\lambda^2+2\mu_p^2 f^2) R_{+}+\frac{i\sqrt{2}\mu_pf^2\partial_r f}
{\lambda-i\sqrt{2}\mu_p f}(f^2\partial_r+g-h)R_+=0
\eqno{(34)}
$$
where $g(r) = {{{f^2 (r)}\over{2r}} + {3\over
    2}{f(r)\partial_r{f(r)}}}$
and $h(r) = {mf(r)\over r}$.
The wave function $R_{-}(r)$ satisfies an equation similar
to that of $R_{+}(r)$ with $h(r)$ replaced by
$-h(r)$.

For $r\rightarrow\infty$, the radial equations
for $R_{+}$ and $R_-$ both reduce to
$$
[\partial_r^2+\frac{5}{r}\partial_r+\frac{4}{r^2}-(\lambda^2+2\mu_p^2f^2)\frac{l^4}{r^4}]R_{\pm}=0,
\eqno{(35)}
$$
keeping leading order terms which gives
$$
\left(\partial_r^2 + \frac{5}{r}\partial_r + \frac{4-2\mu_p^2 l^2}{r^2}\right)R_{\pm}=0.
\eqno{(36)}
$$
Choosing $z=-\frac{1}{4r^4}$, this reduces to
$$
\left(z^2\partial_z^2+\frac{2-\mu_p^2 l^2}{8}\right)R_{\pm}=0
\eqno{(37)}
$$
whose solution is 
$$
R_{\pm} \rightarrow z^{1/2+s},z^{1/2-s},
\eqno{(38)}
$$ 
where $s=\frac{\mu_pl}{2\sqrt{2}}$.
Therefore, the general solutions are $R_+=Az^{1/2+s}+Bz^{1/2-s}$
and $R_-=Cz^{1/2+s}+Dz^{1/2-s}$ when $A,B,C,D$ are arbitrary constants. 
Hence for $r\rightarrow\infty$, $|R_+|$ and $|R_-|$ (as well as $|R_+|^2$ and $|R_-|^2$) both 
decay to zero provided $\mu_p l<\sqrt{2}$. 
It also follows from eqns. (36-38) that asymptotically the first derivatives
of $R_+$ and $R_-$ should reduce to zero 
and thus the asymptotic solutions for $R_+$ and $R_-$ are same with $A=C$
and $B=D$ \cite{isham}.
Now once we know $S_\pm$ and $R_\pm$, we can obtain original
components of spinor by combining them as
$$
\tilde{f}_1=\frac{S_- R_- - S_+ R_+}{2}\rightarrow r^{-2\mp 4s} F\left(a,b,c;\frac{1-x}{2}\right)
\left[\left(\frac{1-x}{2}\right)^{\frac{n+\mu}{2}}\left(\frac{1+x}{2}\right)^{\frac{n-\mu}{2}}\right.
$$
$$
\left.-\left(\frac{1-x}{2}\right)^{\frac{n-\mu}{2}}\left(\frac{1+x}{2}\right)^{\frac{n+\mu}{2}}\right]
\eqno{(39)}
$$
for $r\rightarrow\infty$.
Similarly, other components $f_2$, $g_1$, $\tilde{g}_2$ can be obtained.

\section{Effective potentials of $R_\pm$ fields and comparison with
potentials of similar fields in the Kerr geometry}

\subsection{Effective potentials of $R_\pm$}

\begin{figure}
\begin{center}
\includegraphics[width=10cm,height=14cm,angle=270]{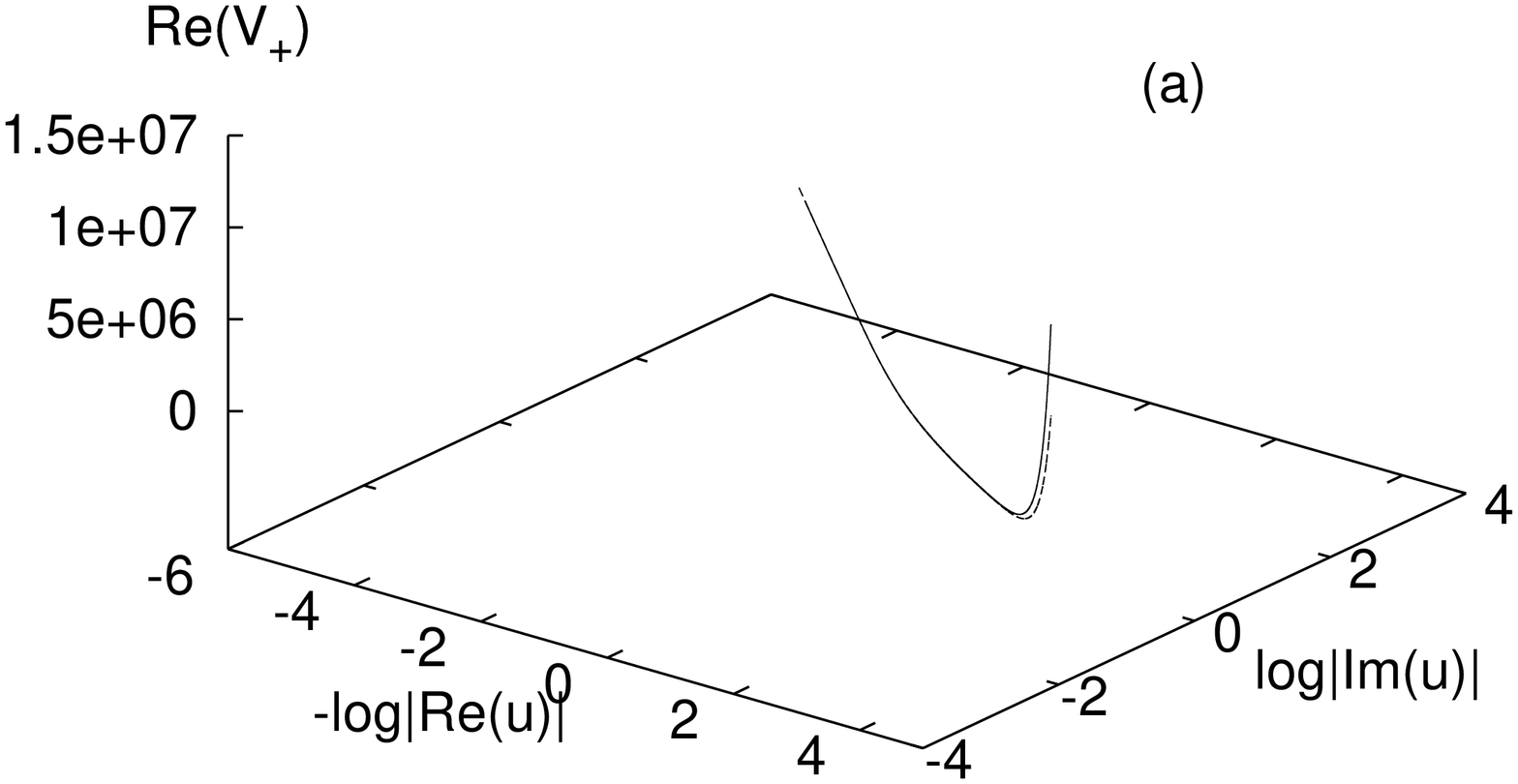}
\includegraphics[width=10cm,height=14cm,angle=270]{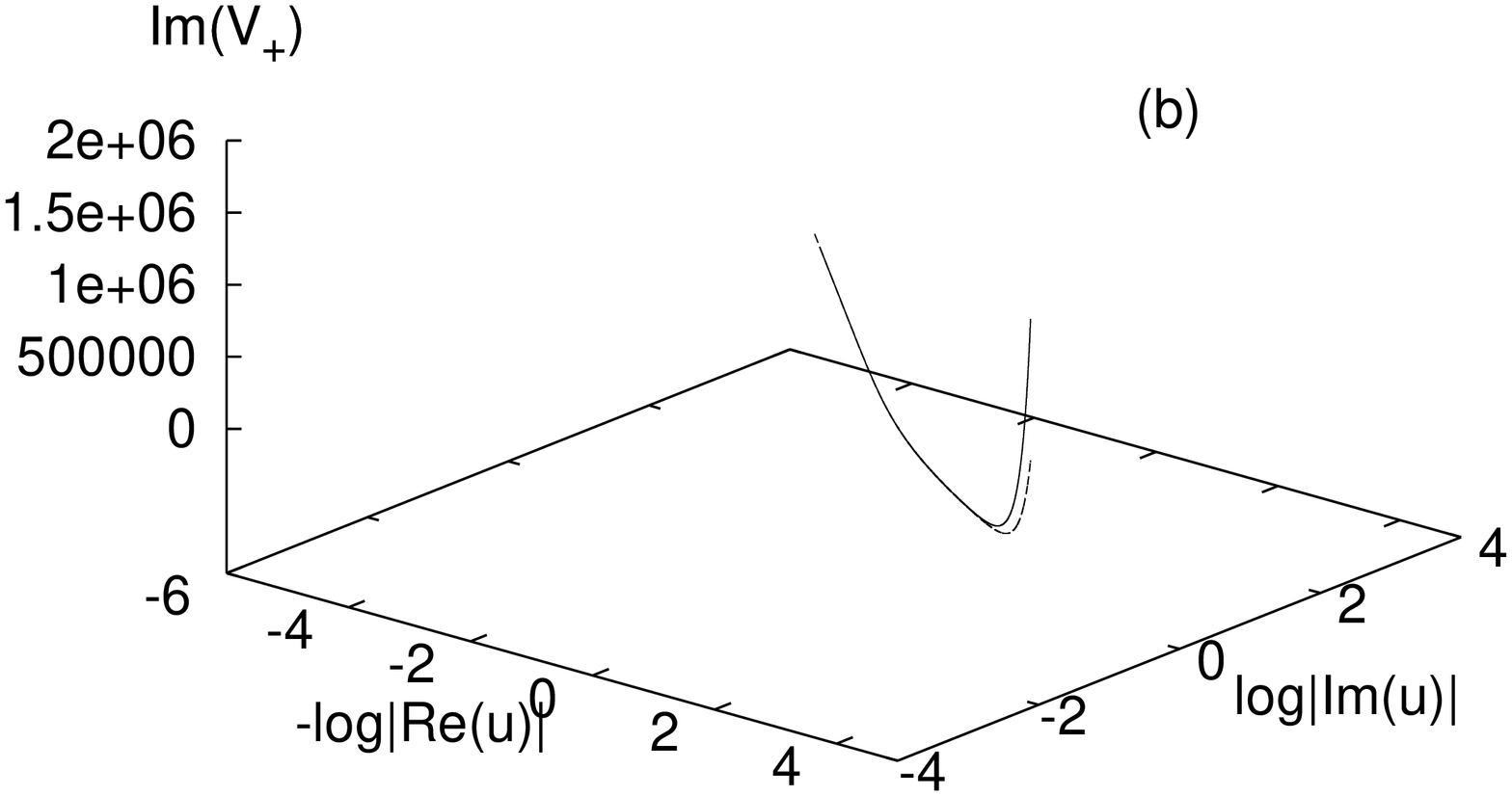}
\end{center}
\caption{Variation of effective potential, (a) real part of $V_+$, (b) imaginary part of $V_+$, 
in the complex plane of $u$ for $l=3$ (solid curve), $1$ 
(dashed curve). Other parameters are $M=1$, $r_h=l$, $m=1/2$, $\mu_p=1$.
} \label{fig1}
\end{figure}

\begin{figure}
\begin{center}
\includegraphics[width=16cm,height=18cm]{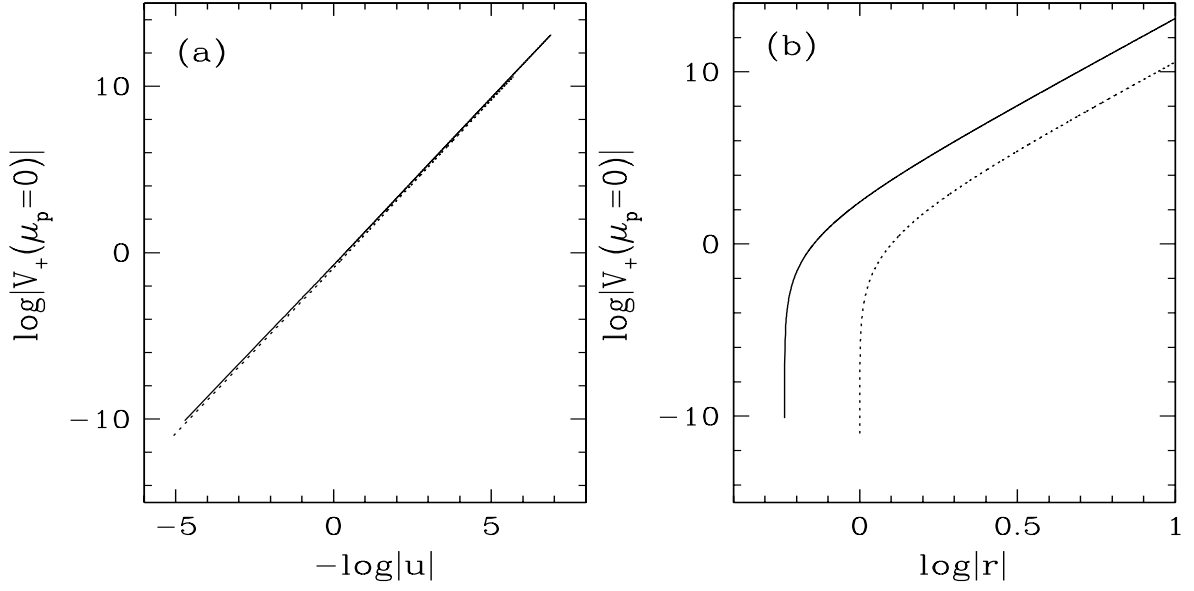}
\end{center}
\caption{Variation of $V_+$ for massless fermion as a function of (a) $u$, 
(b) $r$, for $l=3$ (solid curve), $1$ (dashed curve). 
Other parameters are $M=1$, $r_h=l$, $m=1/2$. 
} \label{fig2}
\end{figure}

Let us define 
$$
f^2 dy=dr
$$
such that the eqn. (34) reduces to
$$
\left[\partial_y^2+\left(2g+\frac{i\sqrt{2}\mu_p(\partial_y f)}{\lambda-
i\sqrt{2}\mu_p f}\right)\partial_y+\left(g^2-h^2+\partial_y(g-h)-
(\lambda^2+2\mu_p^2f^2)+\frac{i\sqrt{2}\mu_p(\partial_y f)}{\lambda-
i\sqrt{2}\mu_p f}(g-h)\right)\right]R_+=0.
\eqno{(40)}
$$
Defining further
$$
u=\int\frac{\lambda-i\sqrt{2}\mu_p f}{r f^3} dy
=\frac{\lambda l^5}{3r_h^5}\left(\frac{3r^2r_h-4r_h^3}{l^3f^3}-3\tan^{-1}\left(\frac{r_h}{lf}\right)\right)
+\frac{i\sqrt{2}\mu_pl^4}{2r_h^4}\left(\frac{r_h^2}{l^2f^2}+2\log\left(\frac{lf}{r}\right)\right)
\eqno{(41)}
$$
eqn. (40) reduces to
$$
\left(\frac{\partial^2}{\partial u^2}+V_+\right)R_+=0
\eqno{(42)}
$$
where the effective potential
$$
V_+=\left(g^2-h^2+\partial_y(g-h)
-(\lambda^2+2\mu_p^2f^2)+\frac{i\sqrt{2}\mu_p(\partial_y f)}{\lambda-
i\sqrt{2}\mu_p f}(g-h)\right)
\frac{r^2 f^6}{(\lambda-i\sqrt{2}\mu_p f)^2}.
\eqno{(43)}
$$
Similarly the potential $V_-$ for $R_-$ can be obtained which is 
same as $V_+$ except $h(r)$ replaced by $-h(r)$. Figure \ref{fig1} describes
behavior of $V_+$ in the complex plane of $u$ for two values of cosmological
constant $\Lambda=-1/l^2$. In Fig. \ref{fig2}, we show $V_+$ for massless fermion. 
It is found that close to the black hole horizon when $Re(u)\rightarrow -\infty$
(and $r\rightarrow r_h$) the potential barrier vanishes, while at far away it
diverges when $Re(u)\rightarrow 0$ (and $r\rightarrow\infty$). This assures
that the radial solution vanishes asymptotically what we indeed show
in eqns. (37)-(39). Behavior of $V_-$ is very similar to $V_+$. Note that for 
all the figures we choose the separation constant $\lambda=1$. In 
principle $\lambda$ should have computed explicitly, especially from the
angular Dirac equation, which might have been different than unity. However,
this does not affect the qualitative feature of $V_\pm$ and thus we keep the exact computation
of $\lambda$ as a future mission.

\subsection{Effective potentials in the Kerr geometry}

First of all let us recall the Dirac equation in Kerr geometry
\cite{chandra83}
$$
{\cal D}_0f_1+2^{-1/2}{\cal L}_{1/2}f_2=(i\mu_*r+a\mu_*\cos\theta)g_1,
\eqno{(44a)}
$$
$$
\Delta{\cal D}_{1/2}^{\dag}f_2-2^{1/2}{\cal L}_{1/2}^{\dag}f_1=
-2(i\mu_*r+a\mu_*\cos\theta)g_2,
\eqno{(44b)}
$$
$$
{\cal D}_0g_2-2^{-1/2}{\cal L}_{1/2}^{\dag}g_1=(i\mu_*r-a\mu_*\cos\theta)f_2,
\eqno{(44c)}
$$
$$
\Delta{\cal D}_{1/2}^{\dag}g_1+2^{1/2}{\cal L}_{1/2}g_2=-2(i\mu_*r-a\mu_*\cos\theta)f_1,
\eqno{(44d)}
$$
where $f_1,f_2,g_1,g_2$ are components of spinor and
$$
{\cal D}_n=\partial_r+\frac{iK}{\Delta}+2n\frac{r-M}{\Delta},
\eqno{(45a)}
$$
$$
{\cal D}_n^{\dag}=\partial_r-\frac{iK}{\Delta}+2n\frac{r-M}{\Delta},
\eqno{(45b)}
$$
$$
{\cal L}_n=\partial_\theta+Q+n {\rm cot}\theta,
\eqno{(46a)}
$$
$$
{\cal L}_n^{\dag}=\partial_\theta-Q+n{\rm cot}\theta,
\eqno{(46b)}
$$
with $K=(r^2+a^2)\sigma+am$, $Q=a\sigma {\rm sin}\theta+m {\rm cosec}\theta$.
We follow the usual convention for various notations (see e.g. \cite{chandra83})
which we do not repeat here.
If one compares this set of equations with the corresponding eqns. (20a-d) 
obtained for the CCBH spacetime, then
it is easy to understand that there are differences in structure of equations. 
While in the CCBH background the angular and radial functions are 
coupled with respectively radial and angular operators, in the Kerr
metric they are completely decoupled due to very nature of the spacetime.
Therefore, after choosing 
$f_1(r, \theta)=R_{-1/2}(r)S_{-1/2}(\theta)$, $f_2(r, \theta)=R_{1/2}(r)S_{1/2}(\theta)$,
$g_1(r, \theta)=R_{1/2}(r)S_{-1/2}(\theta)$, $g_2(r, \theta)=R_{-1/2}(r)S_{1/2}(\theta)$,
the Dirac equation in Kerr geometry  
can easily be separated into radial and angular part given by \cite{chandra76,chandra83,mc}
$$
\Delta^{\frac{1}{2}}{\cal D}_{0} R_{- \frac{1}{2}}
= ( \lh + i m_p r) \Delta^{\frac{1}{2}} {R}_{+ \frac{1}{2}},
\eqno{(47a)}
$$
$$
\Delta^{\frac{1}{2}} {\cal D}_{0}^{\dag} \Delta^{1 \over 2}
{R}_{+{\frac{1}{2}}} = ( \lh - i m_p  r)  {R}_{-{1 \over 2}},
\eqno{(47b)}
$$
$$
{\cal L}_{1 \over 2} S_{+{1 \over 2}} = - (\lh -a m_p \cos \theta) S_{- {1 \over 2}},
\eqno{(48a)}
$$
$$
{\cal L}_{1 \over 2}^{\dag} S_{-{1 \over 2}} = + (\lh+a m_p \cos \theta) S_{+{1\over2}},
\eqno{(48b)}
$$
where $m_p=2^{1/2}\mu_*$ is the mass of the particle, $\lh$ is the separation
constant and $2^{1/2}R_{-1/2}$ is redefined as $R_{-1/2}$.

However, this is not obvious in the CCBH geometry. 
In this case, we change the basis by combining the basic spinors 
and obtain eqns. (25a-d) which are {\it mathematically} analogous to the set of eqns. (44a-d).
Now as $\Psi$-s in eqns. (25a-d) are linear combination of spinors rather than the spinors itself,
we are not able to separate the equations for radial and angular spinors, $R_{\pm 1/2}(r)$ and 
$S_{\pm 1/2}(\theta)$, as could do in the Kerr spacetime. However, we obtain the separated 
radial and angular equations for spinors in new basis given by eqns. (29)-(30).

In the Kerr geometry we change the independent variable $r$ to $r_*$ such that
$$
r_{*} = r + \frac{2M r_+ + am/\sigma} {r_+ - r_-} {\rm log}
\left({r \over r_+} - 1\right) - \frac{2M r_- + am/\sigma} {r_+ - r_-} {\rm log}
\left({r \over r_{-}} - 1\right)
\eqno{(49)}
$$
and choose $R_{-{1\over 2}}=P_{-{1\over 2}}$, $\Delta^{1\over 2}
R_{+ {1 \over 2}} = P_{+ {1 \over 2}}$. Then by defining
$$
(\lh \pm i m_p r) = exp ({\pm i \theta}) \sqrt{{\lh}^{2} + m_p^2 r^2},\,\,
\tan\theta=\frac{m_p r}{\lh},
\eqno{(51)}
$$
$$
P_{\pm {1 \over 2}} = \psi_{\pm {1 \over 2}}\  exp\left[\mp{1 \over 2} i\, \tan^{-1} \left({{m_p r}
\over \lh}\right)\right],
\eqno{(50)}
$$
$$
Z_{\pm} = \psi_{+ {1 \over 2}} \pm \psi_{-{1 \over 2}}
\eqno{(52)}
$$
and combining the eqns. (47a-b) we obtain \cite{chandra83,m99,mc},
$$
\left(\frac{d} {d{\hat r}_*} - W\right) Z_+ = i \sigma Z_-,
\eqno{(53a)}
$$
and
$$
\left(\frac{d} {d{\hat r}_*} + W\right) Z_- = i \sigma Z_+,
\eqno{(53b)}
$$
where,
$$
{\hat{r}}_*=r_*+\frac{1}{2\sigma}\tan^{-1}\left(\frac{m_pr}{\lh}\right),
\eqno{(54})
$$
$$
W = \frac{\Delta^{1 \over 2} (\lh^{2} + m_p^2 r^2)^{3/2}}{\omega^2(\lh^2+m_p^2r^2)
+\lh m_p \Delta/2\sigma},\,\,\,\omega^2=\frac{K}{\sigma}.
\eqno{(55)}
$$
Physically the radial eqns. (53a-b) are similar to 
the set of eqns. (29) obtained for the CCBH metric, except the independent variable $\hat{r}_*$
is Cartesian like while $r$ is the radius vector of polar coordinate system.
Either set is describing the spinor fields 
not in the original basis but in the transformed one like the linear combination of original ones.
Decoupling eqns. (53a-b) we obtain
$$
\left(\frac{d^2} {{d {\hat r}_*}^2} + \sigma^2\right) Z_\pm = V_{k\pm} Z_\pm, 
\eqno{(56)}
$$
where 
$$
V_{k\pm} = W^{2} \pm {dW \over d\hat{r}_{*}}
$$
$$
={{\Delta^{1 \over 2}(\lh^{2} + m_p^{2} r^{2})^{3/2}} \over {[ \omega^{2}(\lh^{2} + m_p^{2}
r^{2}) + \lh m_p \Delta/2 \sigma]^{2}}}[\Delta^{1 \over 2}(\lh^{2} + m_p^{2} r^{2})^{3/2} \pm
 ((r-M)(\lh^{2} + m_p^{2} r^{2}) + 3m_p^{2} r \Delta)]
$$
$$
\mp {{\Delta^{3 \over 2}(\lh^{2} + m_p^{2} r^{2})^{5/2}} \over {[ \omega^{2}(\lh^{2} + m_p^{2}
r^{2}) + \lh m_p \Delta/2 \sigma]^{3}}}[2r(\lh^{2} + m_p^{2} r^{2}) + 2 m_p^{2} \omega^{2} r +
 \lh m_p (r-M)/\sigma] 
\eqno{(57)}
$$
which carries similar information as eqn. (42) does for $R_+$.
Now if we compare eqns. (53a-b), which generate $V_{k+}$ in eqn. (57),  with eqns. (29)
generating $V_+$ in eqn. (42), then the difference is obvious. Hence, the effective
potential $V_\pm$ for $R_\pm$ fields in the CCBH spacetime come out to be different (which
are complex functions) than $V_{k\pm}$ for $Z_\pm$ fields in the Kerr spacetime.
Note that unlike $V_+$ described in Figs. \ref{fig1}, \ref{fig2}, $V_{k\pm}$ attain
a finite value at $\hat{r}_*\rightarrow -\infty$ (close to the black hole horizon) 
and $\infty$ \cite{chandra83,mc}. In addition, by the very nature of spacetime and corresponding 
construction one can obtain the original spinors
$R_{\pm 1/2}$ from $Z_\pm$. However, there is no obvious way to obtain original spinors
in the CCBH spacetime from $R_\pm$, what one needs to define for the sake of
separation, which is understood from eqns. (20), (25), (28) as described
before.   

\section{Summary}

We have considered the spin-half fermion  
field in the $3+1$-dimensional CCBH background.
The Dirac equation is obtained by using the
Newman-Penrose formalism. We show that the Dirac equation is separable
only with the change of basis. The new spinors are linear
combination of original spinors giving rise to the Dirac equation
separable into radial and angular parts. We obtain the corresponding solutions.
We discuss the difference in the Dirac equation in the CCBH spacetime with that
in the Kerr geometry recalling the geometrical conditions to be satisfied
to separate them out \cite{cm}. We show that the structure of equations in the earlier
case is different than the later one (due to difference in structure of the
spacetimes) leading to a different separation and solution procedure.
Now one can consider the Dirac equation solution in the CCBH spacetime to
study the thermodynamical properties of spinor field as was done earlier
for scalar field \cite{K1}.

Note that the coordinate system used in the text
does not cover the entire manifold outside the 
horizon completely. There are other foliations which
cover the outside horizon region completely \cite{HP}. 
However in this foliation the metric becomes
explicitly time dependent. 


\end{document}